\documentclass[12pt]{article}

\usepackage{graphicx}
\begin{document}

\begin{center}
{\bf Black hole solution in the framework of $\arctan$-electrodynamics} \\
\vspace{5mm} S. I. Kruglov
\footnote{E-mail: serguei.krouglov@utoronto.ca}
\underline{}
\vspace{3mm}

\textit{Department of Chemical and Physical Sciences, University of Toronto,\\
3359 Mississauga Road North, Mississauga, Ontario L5L 1C6, Canada} \\
\vspace{5mm}
\end{center}

\begin{abstract}
An $arctan$-electrodynamics coupled with the gravitational field is investigated.
We obtain the regular black hole solution that at $r\rightarrow \infty$ gives corrections to the Reissner-Nordstr\"om solution.
The corrections to Coulomb's law at $r\rightarrow\infty$ are found. We evaluate the mass of the black hole that is a function of the dimensional parameter $\beta$ introduced in the model. The magnetically charged black hole was investigated and we have obtained the magnetic mass of the black hole and the metric function at $r\rightarrow \infty$. The regular black hole solution is obtained at $r\rightarrow 0$ with the de Sitter core. We show that there is no singularity of the Ricci scalar for electrically and magnetically charged black holes.
Restrictions on the electric and magnetic fields are found that follow from the requirement of the absence of superluminal sound speed and the requirement of a classical stability.
\end{abstract}

\section{Introduction}

The self-interaction of photons, due to quantum corrections \cite{Heisenberg}, \cite{Schwinger}, leads to the modification of Maxwell's
electrodynamics which becomes nonlinear electrodynamics (NLED). Thus, in the case of strong electromagnetic fields Maxwell's electrodynamics has to be modified \cite{Jackson}. For weak electromagnetic fields NLED is converted into classical electrodynamics which can be considered as an approximation. Well-known Born-Infeld electrodynamics (BIE) \cite{Born} is an example of NLED that can solve
the problem of singularity and infinite electromagnetic energy of charged point-like particles.
In some models of NLED \cite{Shabad}, \cite{Kruglov}, \cite{Kruglov2} also  problems of singularities and infinite electromagnetic energy are absent. It should be mentioned that NLED coupled to gravitational fields may result in the universe acceleration \cite{Garcia}-\cite{Kruglov4}.
At the same time, electromagnetic fields within BIE can not drive the universe to accelerate \cite{Novello1}. In addition, BIE possesses the problem of causality \cite{Quiros}. Therefore, the study of different models of NLED coupled to gravity is of definite interest.
In this paper we find the solution for electrically and magnetically charged black holes within $arctan$-electrodynamics, that is NLED, proposed in \cite{Kruglov4}.
It was demonstrated in \cite{Ayon} that there is a regular exact black hole solution in General Relativity (GR) coupled with NLED. Authors used a Legendre transformation \cite{Garcia1} to explore so called $P$ frame but not the Lagrangian formalism.
Some black hole solutions, which approach to the Reissner-Nordstr\"{o}m (RN) solution at $r\rightarrow \infty$, within some models of NLED coupled to gravitational fields were found in \cite{Breton} - \cite{Kruglov7}. It was demonstrated in \cite{Bronnikov} that any Lagrangian of NLED coupled to GR for regular electrically charged solutions is branching.

The structure of the paper is as follows. In section 2 we study energy conditions in the model under consideration.
$Arctan$-electrodynamics coupled to gravity is investigated in section 3. We obtain a solution for the electric field and corrections to the Coulomb law in subsection 3.1. The electric-magnetic duality between $P$ and $F$ frames is studied in subsection 3.2. In section 4 we obtain the regular black hole solution, in the framework of electrically charged black hole, which asymptotically approaches to RN solution.
It was shown that there is no singularity of the Ricci scalar.
The mass of the black hole is evaluated and asymptotic metric function is found. Magnetically charged black hole is investigated in section 5.
The regular black hole solution with the de Sitter core is found at $r\rightarrow 0$.
We obtain the magnetic mass of the black hole and metric function at $r\rightarrow \infty$ and $r\rightarrow 0$.
It was shown that there is no singularity of the Ricci curvature at $r\rightarrow \infty$ and $r\rightarrow 0$.
In Sec. 6 we study restrictions that follow from the requirement that the sound speed should be less that the light speed and a classical stability has to take place.
Section 7 is devoted to the results obtained.

We use the units $c=\hbar=1$ and the metric signature $\eta=\mbox{diag}(-1,1,1,1)$.

\section{The model of nonlinear electrodynamics}

We investigate $\arctan$-electrodynamics proposed in \cite{Kruglov4} with the Lagrangian density
\begin{equation}
{\cal L} = -\frac{1}{\beta}\arctan(\beta{\cal F}),
 \label{1}
\end{equation}
where the parameter $\beta$ has the dimension of the (length)$^4$,  ${\cal F}=(1/4)F_{\mu\nu}F^{\mu\nu}=(1/2)(\textbf{B}^2-\textbf{E}^2)$ and $F_{\mu\nu}$ is the strength tensor of electromagnetic fields.
The symmetric energy-momentum tensor is given by \cite{Kruglov4}
\begin{equation}
T^{\mu\nu}=-\frac{F^{\mu\lambda}F^\nu_{~\lambda}}{1+(\beta{\cal F})^2}-g^{\mu\nu}{\cal L},
\label{2}
\end{equation}
possessing the trace
\begin{equation}
{\cal T}\equiv T_{\mu}^{~\mu}=\frac{4}{\beta}\arctan(\beta{\cal F})-\frac{4{\cal F}}{1+(\beta{\cal F})^2}.
\label{3}
\end{equation}
In the case of Maxwell's electrodynamics $\beta\rightarrow 0$ and the energy-momentum tensor becomes traceless.
The dual invariance and scale invariance are broken due to the presence of the dimensional parameter $\beta$ \cite{Kruglov4}.
To guarantee that the energy density is nonnegative for any local observer,
one needs to investigate the weak energy condition (WEC) \cite{Hawking}
\begin{equation}
\rho\geq 0,~~~\rho+p_m\geq 0 ~~~~ (m=1,~2,~3),
\label{4}
\end{equation}
where the $\rho$ is the energy density and $p_m$ are principal pressures. One
finds from Eq. (2) the energy density
\begin{equation}
\rho=T_0^{~0}=\frac{E^2}{1+(\beta {\cal F})^2}+\frac{1}{\beta}\arctan(\beta {\cal F}).
\label{5}
\end{equation}
For the case of pure electric field, $\textbf{B}=0$, the energy density is
\begin{equation}
\rho_E=\frac{E^2}{1+\beta^2 E^4/4}-\frac{1}{\beta}\arctan(\beta E^2/2).
\label{6}
\end{equation}
The plot of the function $\beta\rho_E$ vs $\beta E^2$ is given in Fig. 1.
\begin{figure}[h]
\includegraphics[height=3.0in,width=3.0in]{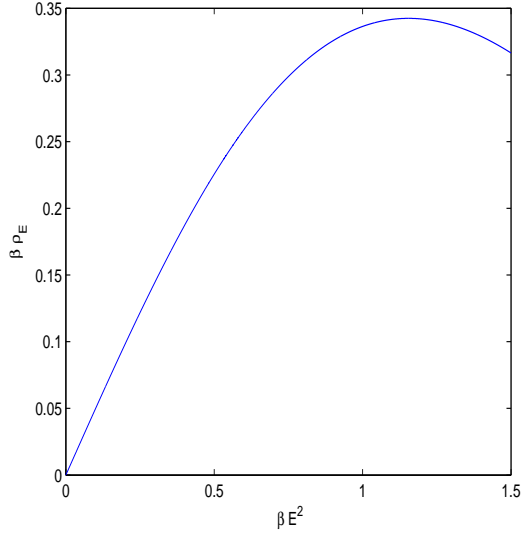}
\caption{\label{fig.1}The function $\beta\rho_E$ vs $\beta E^2$.}
\end{figure}
We will show later that the maximum of the electric field is $E_{max}=(4/3)^{1/4}/\sqrt{\beta} \simeq 1.075/\sqrt{\beta}$, and one can verify that the energy density is positive for the range $0\leq E\leq E_{max}$.
For the case of pure magnetic field, $\textbf{E}=0$, the energy density becomes
\begin{equation}
\rho_M=\frac{1}{\beta}\arctan(\beta B^2/2),
\label{7}
\end{equation}
and it is positive.
The principal pressures are given by
\begin{equation}
p_m=-T_m^{~m}=\frac{B^2-B_mB^m-E_mE^m}{1+(\beta {\cal F})^2}-\frac{1}{\beta}\arctan(\beta {\cal F})~~~~(m=1,~2,~3).
\label{8}
\end{equation}
From Eqs. (5) and (8), one obtains
\begin{equation}
\rho+p_m=\frac{B^2-B_mB^m+E^2-E_mE^m}{1+(\beta {\cal F})^2}\geq 0.
\label{9}
\end{equation}
Thus, WEC is satisfied.
The strong energy condition (SEC) \cite{Hawking} requires relations that hold as follows:
\begin{equation}
\rho+\sum_{m=1}^3p_m\geq 0.
\label{10}
\end{equation}
We obtain from Eqs. (5) and (8)
\begin{equation}
\rho+\sum_{m=1}^3 p_m=\frac{2B^2}{1+(\beta {\cal F})^2}-\frac{2}{\beta}\arctan(\beta {\cal F}).
\label{11}
\end{equation}
It is obvious that for a pure electric field, $\textbf{B}=0$, Eq. (10) is satisfied. For the case of pure magnetic field, SEC is satisfied for
$B<1.66838/\sqrt{\beta}$ as follows from Fig. 2.
\begin{figure}[h]
\includegraphics[height=3.0in,width=3.0in]{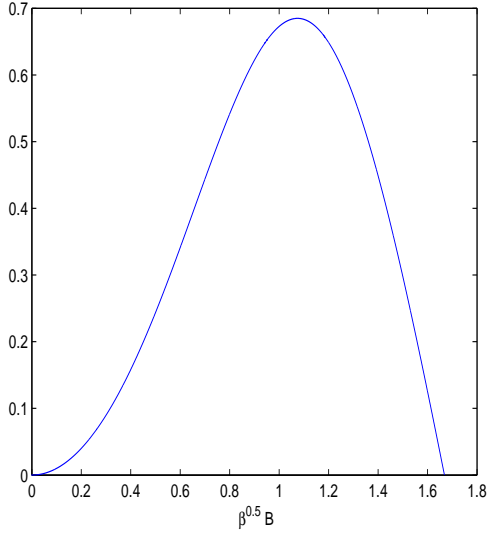}
\caption{\label{fig.2}The function $\beta(\rho_M+\sum p_m)$ vs $\sqrt{\beta} B$.}
\end{figure}
SEC can be represented as
\begin{equation}
\rho +3p\geq 0,
\label{12}
\end{equation}
where the pressure is
\begin{equation}
p={\cal L}+\frac{E^2-2B^2}{3}{\cal L}_{\cal F}=-\frac{1}{\beta}\arctan(\beta {\cal F})+\frac{2B^2-E^2}{3[1+(\beta {\cal F})^2]}=
\frac{1}{3}\sum_{m=1}^3 p_m,
\label{13}
\end{equation}
and ${\cal L}_{\cal F}=\partial {\cal L}/\partial {\cal F}$.
Eq. (12), according to Friedmann's equation, tells us that universe decelerates. Thus, electrically charged universe undergoes the universe deceleration and SEC is satisfied. It should be noted that
magnetically charged universe at $B>1.66838/\sqrt{\beta}$ accelerates \cite{Kruglov4} and SEC is violated.

\section{NLED coupled with GR and black holes}

The action of our model of NLED coupled with GR is given by
\begin{equation}
S=\int d^4x\sqrt{-g}\left[\frac{1}{2\kappa^2}R+ {\cal L}\right],
\label{14}
\end{equation}
where $\kappa^2=8\pi G\equiv M_{Pl}^{-2}$, $G$ is the Newton constant, $M_{Pl}$ is the reduced Planck mass, and $R$ is the Ricci scalar. By varying action (14), one finds the Einstein and NLED equations
\begin{equation}
R_{\mu\nu}-\frac{1}{2}g_{\mu\nu}R=-\kappa^2T_{\mu\nu},
\label{15}
\end{equation}
\begin{equation}
\partial_\mu\left(\frac{\sqrt{-g}F^{\mu\nu}}{1+\left(\beta {\cal F}\right)^2}\right)=0.
\label{16}
\end{equation}
Let us obtain the static charged black hole solutions to Eqs. (15),(16). For this purpose we explore the spherically symmetric line element in
$(3+1)$-dimensional spacetime,
\begin{equation}
ds^2=-f(r)dt^2+\frac{1}{f(r)}dr^2+r^2(d\vartheta^2+\sin^2\vartheta d\phi^2).
\label{17}
\end{equation}

\subsection{Electrically charged black hole}

Now we consider the case $\textbf{B}=0$ for which the vector-potential has non-zero component $A_0(r)$. By virtue of the equality ${\cal F}=-[E(r)]^2/2$, one obtains from (16) the equation as follows:
\begin{equation}
\partial_r\left(\frac{4r^2 E(r)}{4+\beta^2 [E(r)]^4}\right)=0.
\label{18}
\end{equation}
Integrating Eq. (18), we find
\begin{equation}
4r^2E(r)=Q\left(4+\beta^2 [E(r)]^4\right),
\label{19}
\end{equation}
and $Q$ being the integration constant. For a convenience we introduce the dimensionless variables
\begin{equation}
y=\frac{4r^2}{Q\sqrt{\beta}},~~~~x=\sqrt{\beta}E(r).
\label{20}
\end{equation}
From Eqs. (19), (20), we obtain the equation
\begin{equation}
x^4-xy+4=0.
\label{21}
\end{equation}
The plot of the function $y(x)$ is given in Fig. 3.
\begin{figure}[h]
\includegraphics[height=3.0in,width=3.0in]{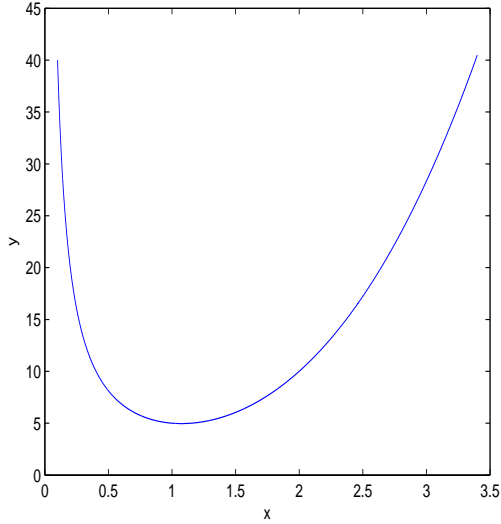}
\caption{\label{fig.3}The function $y(x)$ .}
\end{figure}
It follows from Fig. 3 that there are two branches of the function $E(r)$. The physical branch corresponds to the electric field
which decreases with the $r$ and for the nonphysical branch the $E(r)$ increases with the $r$.
The function $y(x)$ possesses a minimum at $x=(4/3)^{1/4}$ that corresponds to $E=[4/(3\beta^2)]^{1/4})$ and
$r_{min}=(4/3)^{3/8}\beta^{1/4}\sqrt{Q}$. As a result, Eq. (21) has the real solutions if $r\geq r_{min}$.
Therefore, there is no singularity of the electric field $E$.
Thus, the maximum electric field is given by $E_{max}=[4/(3\beta^2)]^{1/4})\simeq 1.075/\sqrt{\beta}$ and the electric field decreases with $r$. The real solution to Eq. (21) (for $x\leq(4/3)^{1/4}$, $y>0$) is given by
\[
x=\sqrt{\frac{2}{\sqrt{3}}\cosh\left(\frac{\varphi}{3}\right)}+\sqrt{\frac{y3^{1/4}}{4\sqrt{2\cosh\left(\frac{\varphi}{3}\right)}}-
\frac{2}{\sqrt{3}}\cosh\left(\frac{\varphi}{3}\right)},
\]
\begin{equation}
\varphi =\ln\left(\frac{3\sqrt{3}y^2}{128}+\sqrt{\frac{27y^4}{128^2}-1}\right).
\label{22}
\end{equation}
Solution (22) at $y\rightarrow \infty$ leads to $x\rightarrow 0$.
The asymptotic of the function $x(y)$ (22) at $y\rightarrow \infty$ is
\begin{equation}
x=\frac{4}{y}+\frac{1024}{27y^{11/3}}+\frac{4352}{81y^{5}}+\frac{262144}{243y^{19/3}}+{\cal O}(x^{-20/3}).
\label{23}
\end{equation}
From Eqs. (20) and (23) we find the asymptotic value of the electric field at $r\rightarrow \infty$
\begin{equation}
E(r)=\frac{Q}{r^2}+\beta\left(\frac{4^{4/3}\beta^{1/3} Q^{11/3}}{27r^{22/3}}+\frac{17\beta Q^5}{32r^{10}}+\frac{4^{8/3}\beta^{5/3} Q^{19/3}}{243r^{38/3}}+{\cal O}(r^{-40/3})\right).
\label{24}
\end{equation}
It follows from Eq. (24) that the integration constant $Q$ is the charge. The first term in the right side of Eq. (24) is Coulomb's law and next terms give the corrections to Coulomb's law at $r\rightarrow\infty$. At $\beta=0$ the Coulomb law $E=Q/r^2$ is recovered.

The causality principle says that the group velocity of elementary excitations over a background should be less than light speed \cite{Shabad1}.
This gives the restriction ${\cal L}_{\cal F}<0$ that is satisfied for any electric and magnetic fields as ${\cal L}_{\cal F}=-1/[1+(\beta{\cal F})^2]$. The unitarity principle requires ${\cal L}_{\cal F}+2{\cal F}{\cal L}_{{\cal F}{\cal F}}\leq 0$ \cite{Shabad1} and guarantees the positive definiteness of the norm of elementary excitations of the vacuum. We find from Eq. (1)
\begin{equation}
{\cal L}_{\cal F}+2{\cal F}{\cal L}_{{\cal F}{\cal F}}=\frac{3(\beta {\cal F})^2-1}{[1+(\beta {\cal F})^2]^2}.
\label{25}
\end{equation}
For pure electric field ($\textbf{B}=0$) the unitarity principle gives $E\leq E_{max}=\sqrt{2}/(3^{1/4}\sqrt{\beta})$, that is satisfied.
It follows from Eq. (25) that for pure magnetic field ($\textbf{E}=0$) the unitarity principle requires that $B\leq \sqrt{2}/(3^{1/4}\sqrt{\beta})\simeq  1.075/\sqrt{\beta}$.

For a system with the spherical symmetry the equation $\rho=T^0_{~0}=T^r_{~r}=-p_r$ holds so that $\rho+p_r=0$. The tangential pressure is $p_\perp =-T^\vartheta_{~\vartheta} =-T^\phi_{~\phi}$ \cite{Tolman} and $p_\perp =-\rho-r\rho'/2$ ($\rho'=d\rho/dr$) \cite{Dymnikova}.
For the case $\textbf{B}=0$, $\textbf{E}\neq 0$ we obtain from Eq. (6)
\[
\rho_E'(r)=\frac{4EE'(4-3\beta^2 E^4)}{(4+\beta^2 E^4)^2}.
\]
As $E\leq E_{max}$ the sign of $\rho_E'(r)$ depends on the sign of $E'(r)$. It follows from Fig. 3 that for the physical branch $E'(r)\leq 0$
and $\rho_E'(r)<0$, so that $p_\perp +\rho_E \geq 0$ and WEC is satisfied.

\subsection{Electric-magnetic duality}

In some models of NLED there is a duality between ${\cal F}$ framework and
${\cal P}$ framework obtained by a Legendre transformation [17]. This means that any spherically symmetric
solution in the framework of the Lagrangian $L({\cal F})$ has a counterpart in the ${\cal P}$ framework by the
substitution ${\cal F}\rightarrow {\cal P}$, ${\cal L}\rightarrow {\cal H}$, $F_{01}\rightarrow P_{23}$, $F_{23}\rightarrow P_{01}$ and conversely. Thus, the duality connects solutions in different theories. The reason for the study of the ${\cal P}$ framework is to investigate another theory with the Lagrangian density ${\cal H}$ and automatically obtain the magnetic solution from the electric solution in the original theory with the Lagrangian density ${\cal L}$. But this is possible only in the case if a duality between ${\cal F}$ framework and ${\cal P}$ framework holds.
So, one can make a Legendre transformation \cite{Garcia1} and comes to the $P$ framework, i.e. to another form of NLED. Let us introduce the tensor
\begin{equation}
P_{\mu\nu}=\frac{1}{2}F_{\mu\nu}{\cal L}_{\cal F} =-\frac{F_{\mu\nu}}{2\left[1+(\beta{\cal F})^2 \right]}.
\label{26}
\end{equation}
From Eq. (26) we obtain the invariant
\begin{equation}
P=P_{\mu\nu}P^{\mu\nu}=\frac{{\cal F}}{\left[1+(\beta{\cal F})^2 \right]^2}.
\label{27}
\end{equation}
One may introduce the Hamilton-like variable
 \begin{equation}
{\cal H}=2{\cal F}{\cal L}_{\cal F}-{\cal L}=\frac{1}{\beta}\arctan(\beta{\cal F})-\frac{{2\cal F}}{\left[1+(\beta{\cal F})^2 \right]}.
\label{28}
\end{equation}
We find from Eqs. (28) and (5) that ${\cal H}=\rho_E$ (at $\textbf{B}=0$) is the energy density. One can verify that he relations
\begin{equation}
{\cal L}_{\cal F}{\cal H}_P=1,~~~~P{\cal H}_P^2={\cal F},~~~~{\cal L}=2P{\cal H}_P-{\cal H}
\label{29}
\end{equation}
hold with
\begin{equation}
{\cal H}_P=\frac{\partial {\cal H}}{\partial P}=-\left[1+(\beta{\cal F})^2 \right].
\label{30}
\end{equation}
As follows from the plot of the function $P({\cal F})$, represented in Fig. 4, the function ${\cal F}(P)$ is not a monotonic function.
\begin{figure}[h]
\includegraphics[height=3.0in,width=3.0in]{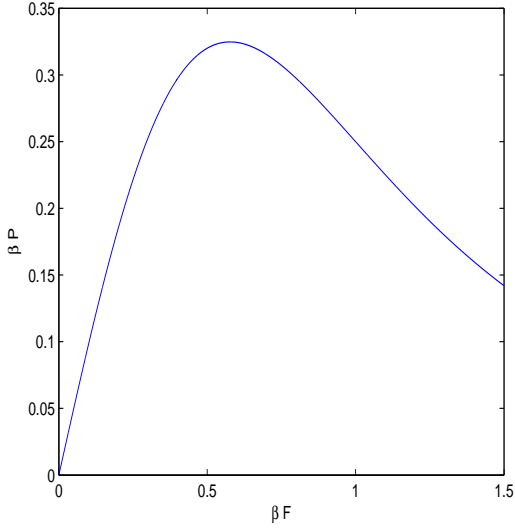}
\caption{\label{fig.4}The function $\beta P$ vs $\beta {\cal F}$.}
\end{figure}
As a result, according to the criterium of Bronnikov \cite{Bronnikov}, there is no a one to one correspondence between two frames, ${\cal F}$ and $P$. In other words, the electric-magnetic duality between ${\cal F}$ and $P$ frames is broken. Thus, a regular electric solution in the $P$ framework corresponds to different Lagrangians in different spacetimes \cite{Bronnikov}.
For weak fields, $\beta{\cal F}\ll 1$, the model with the Lagrangian density (1) and the model with the Hamiltonian-like quantity (28) are converted into Maxwell's theory, ${\cal L}={\cal H}=-{\cal F}$.

\section{Asymptotic Reissner-Nordstr\"{o}m black hole solution}

The expression for Ricci scalar follows from Eq. (15),
\begin{equation}
R=\kappa^2{\cal T}.
\label{31}
\end{equation}
From Eqs. (3) and (31) at $\textbf{B}=0$, we obtain the Ricci curvature
\begin{equation}
R=\kappa^2\left(\frac{2E^2}{1+\beta^2E^4/4}-\frac{4}{\beta}\arctan\left(\beta E^2/2\right)\right).
\label{32}
\end{equation}
According to Eq. (24) the electric field at $r\rightarrow \infty$ goes to zero, $E(r)\rightarrow 0$, and therefore, the Ricci scalar
also at $r\rightarrow \infty$ approaches to zero, $R \rightarrow 0$. At infinity spacetime becomes Minkowski's spacetime (flat).
Using Eqs. (24) and (32) at $r\rightarrow\infty$, we find the Ricci scalar asymptotic
\begin{equation}
R=-\kappa^2\beta^2\left(\frac{Q^6}{3r^{12}}+\frac{2^{11/3}\beta^{4/3} Q^{26/3}}{27r^{52/3}}+ \frac{2\beta^2 Q^{10}}{405r^{20}} +{\cal O}(r^{-68/3})\right).
\label{33}
\end{equation}
It should be mentioned that there is no singularity of the Ricci scalar (32) due to the existence of the minimal length
$r_{min}=(4/3)^{3/8}\beta^{1/4}\sqrt{Q}$ at which the maximum of the electric field is $E_{max}=[4/(3\beta^2)]^{1/4}$.
The metric function $f(r)$ and the mass function $M(r)$ are defined as
\begin{equation}
f(r)=1-\frac{2GM(r)}{r},~~~~M(r)=\int_{r_{min}}^r\rho_E(r)r^2dr=m-\int^\infty_r\rho_E(r)r^2dr.
\label{34}
\end{equation}
The mass of the black hole is $m=\int_{r_{min}}^\infty\rho_E(r)r^2dr$, where $r_{min}=(4/3)^{3/8}\beta^{1/4}\sqrt{Q}$. With the help of Eqs. (6) and (20) we obtain the mass function
\begin{equation}
M(r)=\frac{Q^{3/2}}{16\beta^{1/4}}\int_{x_{max}}^x\left[\frac{4(3x^4-4)}{\sqrt{x(x^4+4)}}-\frac{(3x^4-4)\sqrt{x^4+4}}{x^{5/2}}
\arctan\left(\frac{x^2}{2}\right)\right] dx,
\label{35}
\end{equation}
where $x_{max}=(4/3)^{1/4}$. From Eq. (35) one finds the mass of the black hole
\begin{equation}
m=M(r=\infty)\simeq\frac{0.451 Q^{3/2}}{\beta^{1/4}}.
\label{36}
\end{equation}
 The black hole mass possesses the electromagnetic origin.
From Eqs. (23) and (35) one can find the asymptotic of the mass function at $r\rightarrow \infty$
\begin{equation}
M(r)=m-\frac{ Q^{2}}{2r}-\frac{ 4^{4/3}\beta^{4/3}Q^{14/3}}{171r^{19/3}}+{\cal O}(r^{-9}).
\label{37}
\end{equation}
We obtain from Eqs. (34) and(37) the metric function
\begin{equation}
f(r)=1-\frac{2Gm}{r}+\frac{G Q^{2}}{r^2}+\frac{ 2^{11/3}\beta^{4/3}GQ^{14/3}}{171r^{22/3}}+{\cal O}(r^{-10}).
\label{38}
\end{equation}
At $\beta=0$  one comes to Maxwell's electrodynamics and solution (38) becomes the RN solution.
At $r\rightarrow\infty$ the spacetime is asymptotically flat. The last terms in Eq. (38) give corrections to RN solution.
Different models of NLED \cite{Breton}, \cite{Hendi}, \cite{Kruglov6} give asymptotic of black hole solution with some corrections. As a result, corrections to the RN solution influence on the event horizon. Thus, the equation
$f(r)=0$ for different parameters of the model can have one root corresponding to the extremal black hole or two roots for the event and Cauchy horizons or there can be no roots corresponding to the naked singularity. The equation $f(r)=0$ for our case is complicated to obtain roots analytically.

\section{Magnetically charged black hole}

In this case, there is only one nonzero component of the field strength tensor in spherical system, $F_{23}=- F_{32}=q\sin\vartheta$ \cite{Bronnikov}, where $q$ is a magnetic charge and the invariant is ${\cal F}=(1/4)F_{\mu\nu}F^{\mu\nu}=q^2/(2r^4)$. Then the energy density
found from (7) becomes
\begin{equation}
\rho_M=-{\cal L} = \frac{1}{\beta}\arctan(\beta{\cal F})=\frac{1}{\beta}\arctan\left(\frac{\beta q^2}{2r^4}\right).
 \label{39}
\end{equation}
By virtue of Eq. (39) we obtain the mass of the black hole
\begin{equation}
m=\int^\infty_0 \rho_M r^2dr=\frac{\pi q^{3/2}\csc(\pi/8)}{6\cdot2^{3/4}\beta^{1/4}} \simeq\frac{0.81355 q^{3/2}}{\beta^{1/4}}.
\label{40}
\end{equation}
The mass function is given by
\begin{equation}
M(r)=\frac{1}{\beta}\int_0^r r^2\arctan\left(\frac{\beta q^2}{2r^4}\right)dr.
\label{41}
\end{equation}
The integral (41) is complicated, and therefore, we obtain its asymptotic.
From Eq. (41) we find the asymptotic of the mass function at $r\rightarrow\infty$
\begin{equation}
M(r)=m-\frac{q^2}{2r} +\frac{\beta^2 q^6}{216r^9}+{\cal O}(r^{-12}).
\label{42}
\end{equation}
One obtains the metric function from Eqs. (34) and (42)
\begin{equation}
f(r)=1-\frac{2Gm}{r}+\frac{Gq^2}{r^2}-\frac{G\beta^2 q^6}{108r^{10}}+{\cal O}(r^{-13}).
\label{43}
\end {equation}
which approaches Minkowski's spacetime at $r\rightarrow \infty$.
At $r\rightarrow 0$ we use the Taylor series
of the $r^2\rho_M$ at $r\rightarrow 0$
\begin{equation}
r^2\frac{1}{\beta}\arctan\left(\frac{\beta q^2}{2r^4}\right)=\frac{\pi r^2}{2\beta}-\frac{2r^6}{q^2\beta^2}+{\cal O}(r^{14}).
\label{44}
\end{equation}
Then one obtains the asymptotic of the mass function at $r\rightarrow 0$
\begin{equation}
M(r)=\frac{\pi r^3}{6\beta}-\frac{2r^7}{7q^2\beta^2}+{\cal O}(r^{15}),
\label{45}
\end{equation}
and the metric function
\begin{equation}
f(r)=1-\frac{\pi Gr^2}{3\beta}+\frac{4Gr^6}{7q^2\beta^2}+{\cal O}(r^{14}).
\label{46}
\end{equation}
It follows from Eq. (46) that we have a regular black hole with the de Sitter core and cosmological constant $\Lambda=\pi G/\beta$ \cite{Dymnikova}, \cite{Hayward}.
From Eqs. (3) and (31) we find the  Ricci scalar
\begin{equation}
R=\kappa^2\left[\frac{4}{\beta}\arctan\left(\frac{\beta q^2}{2r^4}\right)-\frac{8q^2r^4}{4r^8+\beta^2q^4}\right].
\label{47}
\end{equation}
and its asymptotic at $r\rightarrow \infty$
\begin{equation}
R=\kappa^2\left(\frac{\beta^2q^6}{3r^{12}}-\frac{\beta^{4} q^{10}}{10r^{20}} +{\cal O}(r^{-28})\right).
\label{48}
\end{equation}
From Eq. (47) one obtains the regular asymptotic of the  Ricci curvature at $r\rightarrow 0$
\begin{equation}
R=\kappa^2\left(\frac{2\pi}{\beta} - \frac{4r^4}{\beta^2 q^2}+\frac{32r^{12}}{3\beta^4q^6} +{\cal O}(r^{20})\right).
\label{49}
\end{equation}
Thus, there is no singularaty of Ricci's scalar at $r\rightarrow \infty$ and $r\rightarrow 0$. We can not put $\beta=0$ in Eq. (49) because in Maxwell' electrodynamics the trace of electromagnetic fields is zero and the Ricci scalar, according to Eq. (31), is zero.

For the case $\textbf{E}=0$, $\textbf{B}\neq 0$ we have
\[
\rho_M'(r)=\frac{\beta B B'}{1+(\beta B^2/2)^2},
\]
and $B'(r)=-2q/r^3<0$. Thus, $\rho_M'(r)<0$ and $\rho_\bot+\rho_M=-r\rho_M'/2>0$, and as a result, WEC is satisfied.

\section{Sound speed and causality}

The speed of the sound has to be less than the light speed and that gives the restriction $c_s\leq 1$ \cite{Quiros}. A classical stability requires: the square sound speed should be positive, $c^2_s> 0$. The square sound speed is defined as
\begin{equation}
c^2_s=\frac{dp}{d\rho}=\frac{dp/d{\cal F}}{d\rho/d{\cal F}}.
\label{50}
\end{equation}
Let us consider two cases.

1). The black hole is electrically charged, $E\neq 0$, $B=0$.

In this case we obtain  from Eqs. (6) and (13) the square sound speed,
\begin{equation}
c^2_s=\frac{1+5(\beta^2 E^4/4)}{3(1-3\beta^2 E^4/4)}.
\label{51}
\end{equation}
It follows from Eq. (51) that a classical stability, $c^2_s> 0$, leads to $E<E_{max}=(4/3)^{1/4}/\sqrt{\beta}$, that is satisfied.
The requirement that there are not superluminal fluctuations, $c_s^2\leq 1$, gives the restriction
\begin{equation}
E\leq E_c\equiv \frac{\sqrt{2}}{7^{1/4}\sqrt{\beta}}\simeq \frac{0.869}{\sqrt{\beta}}.
\label{52}
\end{equation}
It should be mentioned that $E_c<E_{max}\simeq 1.075/\sqrt{\beta}$, i.e. the electric field should be less than the maximum of the electric field. So, a bound (52) guarantees that the sound speed is less than the speed of light. We note that the analysis of causality in subsection 3.1, based on the approach of the work \cite{Shabad1}, did not give the restriction on the electric field.

2). The black hole is magnetically charged, $E=0$, $B\neq 0$.

From Eqs. (7), (50) and (13) we find the square sound speed
\begin{equation}
c^2_s=\frac{1-7(\beta^2 B^4/4)}{3(1+\beta^2 B^4/4)}.
\label{53}
\end{equation}
Then from Eq. (53) we make a conclusion that a classical stability, $c^2_s> 0$, occurs at $B<\sqrt{2/(\sqrt{7}\beta)}$.
This gives the bound on the distance from the center of the black hole
\begin{equation}
r>\left(\frac{\sqrt{7}\beta q^2}{2}\right)^{1/4}\simeq 1.07 \sqrt{q}\beta^{1/4}.
\label{54}
\end{equation}
It follows from Eq. (53) that the requirement  $c_s^2\leq 1$ (there are not superluminal fluctuations) does not give any restrictions on the magnetic field.

\section{Conclusion}

It was shown that in our $arctan$-model of NLED the WEC and SEC are satisfied. Therefore, the $\arctan$-electrodynamics can be considered as a viable NLED model.
We have obtained the exact dependance of the electric field of charged objects on the distance and
the corrections to Coulomb's law. The electric-magnetic duality between $P$ and $F$ frames is broken in our model.
As a result, it is impossible to obtain the magnetic solution in the model with the Lagrangian density ${\cal H}$
by the substitutions ${\cal F}\rightarrow {\cal P}$, ${\cal L}\rightarrow {\cal H}$, $F_{01}\rightarrow P_{23}$, $F_{23}\rightarrow P_{01}$
 from the electric solution in the theory with the Lagrangian density ${\cal L}$.
NLED coupled to gravitational field was analyzed and we calculated the asymptotic of the Ricci scalar at $r\rightarrow \infty$.
We have demonstrated that there is no singularity of the Ricci scalar because the minimal length
$r_{min}=(4/3)^{3/8}\beta^{1/4}\sqrt{Q}$ exists and the maximum of the electric field is $E_{max}=[4/(3\beta^2)]^{1/4}$.
We calculated the mass of the black hole for electrically and magnetically charged black holes.
The asymptotic of the metric function and corrections to the Reissner-Nordstr\"om solution at $r\rightarrow\infty$ were found. We have investigated magnetically charged black hole and have obtained the magnetic mass of the black hole and the metric function at $r\rightarrow \infty$ and $r\rightarrow 0$. We have shown that at $r\rightarrow 0$ there is a de Sitter core corresponding to the cosmological constant $\Lambda=\pi G/\beta$.
It was demonstrated that the black hole is regular and there is not singularity of the Ricci curvature at $r\rightarrow \infty$ and $r\rightarrow 0$ for magnetically charged black hole.
We found restrictions on the electric and magnetic fields that follow from the requirement of the absence of superluminal sound propagation and the requirement of a classical stability. Thus, there are not superluminal fluctuations if $E<E_c\equiv \sqrt{2}/(7^{1/4}\sqrt{\beta})$ but classical stability takes place for any value of the electric field $E< E_{max}$. For magnetically charged black hole a classical stability occurs at $B<\sqrt{2}/(7^{1/4}\sqrt{\beta})$ that corresponds to $r>(\sqrt{7}\beta q^2/2)^{1/4}$ and there are not superluminal fluctuations at any value of the magnetic field.


\begin{thebibliography}{99}

\bibitem{Heisenberg} W. Heisenberg and H. Euler, Z. Physik, \textbf{98}, 714 (1936) (arXiv:physics/0605038).

\bibitem{Schwinger} J. Schwinger, Phys. Rev. \textbf{82}, 664 (1951).

\bibitem{Jackson} J. D. Jackson, \textit{Classical Electrodynamics, Second Ed.}, John Wiley and Sons, 1975.

\bibitem{Born} M. Born and L. Infeld, Proc. Royal Soc. (London) A \textbf{144}, 425 (1934).

\bibitem{Shabad} D. M. Gitman, A. E. Shabad, Eur. Phys. J. C \textbf{74}, 3186 (2014).

\bibitem{Kruglov} S. I. Kruglov, Ann. Phys. \textbf{353}, 299 (2015) (arXiv:1410.0351).

\bibitem{Kruglov2} S. I. Kruglov, Ann. Phys. (Berlin) \textbf{527}, 397 (2015) (arXiv:1410.7633).

\bibitem{Garcia} R. Garc\'{i}a-Salcedo and N. Breton, Int. J. Mod. Phys. A \textbf{15}, 4341 (2000) (arXiv:gr-qc/0004017).

\bibitem{Camara} C. S. Camara, M. R. de Garcia Maia, J. C. Carvalho and J. A. S. Lima, Phys. Rev. D
    \textbf{69}, 123504 (2004) (arXiv:astro-ph/0402311).


\bibitem{Novello} M. Novello, S. E. Perez Bergliaffa and J. M. Salim, Phys. Rev. D \textbf{69}, 127301
    (2004) (arXiv:astro-ph/0312093).

\bibitem{Novello1} M. Novello, E. Goulart, J. M. Salim and S. E. Perez Bergliaffa, Class. Quant. Grav.
    \textbf{24}, 3021 (2007) (arXiv:gr-qc/0610043).

\bibitem{Vollick} D. N. Vollick, Phys. Rev. D \textbf{78}, 063524 (2008) (arXiv:0807.0448).

\bibitem{Kruglov3} S. I. Kruglov, Phys. Rev. D \textbf{92}, 123523 (2015) (arXiv:1601.06309 [gr-qc]).


\bibitem{Kruglov4} S. I. Kruglov, Int. J. Mod. Phys. D \textbf{25}, 1640002 (2016) (arXiv:1603.07326 [gr-qc]).

\bibitem{Quiros} R. Garc\'{i}a-Salcedo, T. Gonzalez and I. Quiros, Phys. Rev. D \textbf{89}, 084047 (2014) (arXiv:1312.3163)[gr-qc]).

\bibitem{Ayon} E. Ay\'{o}n-Beato and A. Garc\'{i}a, Phys. Rev. Lett. \textbf{80}, 5056 (1998) (arXiv:gr-qc/9911046).

\bibitem{Garcia1} H. Salazar, A. Garc\'{i}a, and J. Pleba\'{n}ski, J. Math. Phys. \textbf{28}, 2171 (1987).

\bibitem{Breton} N. Breton, Phys. Rev. D \textbf{67}, 124004 (2003) (arXiv:hep-th/0301254).

\bibitem{Dymnikova} I. Dymnikova, Clas. Quant. Grav. \textbf{21}, 4417 (2004) (arXiv:gr-qc/0407072).

\bibitem{Hendi} S. H. Hendi, Ann. Phys. \textbf{333}, 282 (2013) (arXiv:1405.5359 [gr-qc]).

\bibitem{Balard} L. Balart and E. C. Vagenas, Phys. Rev. D \textbf{90}, 124045 (2014) (arXiv:1408.0306 [gr-qc]).

\bibitem{Kruglov5} S. I. Kruglov, Int. J. Geom. Meth. Mod. Phys. \textbf{12}, 1550073 (2015) (arXiv:1504.03941).

\bibitem{Kruglov6} S. I. Kruglov, Ann. Phys. (Berlin)  \textbf{528}, 588 (2016) (arXiv:1607.07726) [gr-qc]).

\bibitem{Kruglov7} S. I. Kruglov, Phys. Rev. D \textbf{94}, 044026 (20016) (arXiv:1608.04275 [gr-qc]).

\bibitem{Bronnikov} K. A. Bronnikov, Phys. Rev. D \textbf{63}, 044005 (2001).

\bibitem{Hawking} S. W. Hawking and G. F. R. Ellis, \textit{The Large Scale Structure of Space-Time}, Cambridge Univ. Press, 1973.

\bibitem{Shabad1} A. E. Shabad and V. V. Usov, \textit{Phys. Rev. D} \textbf{83}, 105006 (2011) (arXiv:1101.2343).

\bibitem{Tolman} R. C. Tolman, \textit{Relativity, Thermodynamics and Cosmology}, Clarendon Press, Oxford, 1969.

\bibitem{Hayward} S. A. Hayward, Phys. Rev. Lett. \textbf{96}, 031103 (2006) (arXiv:gr-qc/0506126) [gr-qc]).

\end{thebibliography}
\end{document}